%% file: checking_and_enforcing_security_through_opacity_in_healthcare_applications.tex
\documentclass{llncs}
\usepackage[numbers]{natbib}
\makeatletter
\renewcommand\bibsection%
{
  \section*{\refname
    \@mkboth{\MakeUppercase{\refname}}{\MakeUppercase{\refname}}}
}
\makeatother
\usepackage{etex}
\usepackage{graphics,rotating}
\usepackage{graphics}
\usepackage{dsfont}
\usepackage{pst-grad,pst-node,pstricks,tikz}
\usetikzlibrary{fit}
\usepackage{pgflibraryshapes}
\usepackage{pgfplots}
\usepackage{subfigure}
\usepackage{tabularx}
\usetikzlibrary{shapes,automata,arrows,petri}
\tikzset{initial text={}}

\usepackage{amsfonts}
\usepackage{amsmath, amssymb}

\usepackage[linesnumbered,ruled]{algorithm2e}
\makeatletter
\newcommand{\nosemic}{\renewcommand{\@endalgocfline}{\relax}}% Drop semi-colon ;
\newcommand{\dosemic}{\renewcommand{\@endalgocfline}{\algocf@endline}}% Reinstate semi-colon ;
\let\oldnl\nl% Store \nl in \oldnl
\newcommand{\nonl}{\renewcommand{\nl}{\let\nl\oldnl}}% Remove line number for one line
\makeatother
\usepackage{paralist}
\usepackage{graphics,rotating} %,booktabs}
\usepackage{bigstrut}
\usepackage{multirow}
\usepackage{color, colortbl}
\usepackage{float}	

\definecolor{Gray}{gray}{0.85}
\begin{document}
\title{Checking and Enforcing Security through Opacity in Healthcare Applications}

\author{Rym Zrelli\inst{1} 
 \and Moez Yeddes\inst{2} \and Nejib Ben Hadj-Alouane\inst{1} }
\authorrunning{R. Zrelli et al.} % abbreviated author list (for running head)
%\tocauthor{Amina Bourouis, Kais Klai, Nejib Ben Hadj-Alouane}
%
\institute{OASIS Reasearch Lab (ENIT), University of Tunis El Manar, Tunis, Tunisia
\and
OASIS Reasearch Lab (INSAT), University of Carthage, Tunis, Tunisia}

\maketitle

\begin{abstract}
The Internet of Things (IoT) is a paradigm that can tremendously revolutionize health care thus benefiting both hospitals, doctors and patients. In this context, protecting the IoT in health care against interference, including service attacks and malwares, is challenging. Opacity is a confidentiality property capturing a system's ability to keep a subset of its behavior hidden from passive observers. In this work, we seek to introduce an IoT-based heart attack detection system, that could be life-saving for patients without risking their need for privacy through the verification and enforcement of opacity. Our main contributions are the use of a tool to verify opacity in three of its forms, so as to detect privacy leaks in our system. Furthermore, we develop an efficient, Symbolic Observation Graph (SOG)-based algorithm for enforcing opacity.
\end{abstract}

%begin article
\input{intro}
\input{preli}
\input{case}

\input{verif}
\input{app}

\input{conc}


%bibliography

\input{ISYCC-2017.bbl}
\bibliographystyle{splncs03}
%\bibliography{mybib}{}
\clearpage
%\input{computing}
%\input{proof}
%\input{apen}
\end{document}

%% file: intro.tex
\addcontentsline{toc}{section}{Introduction}
\section{Introduction}

Real-world usage of IoT in health-care necessitates the dealing with new security challenges. In fact, and since this type of application would handle medical and personal information, their employment carries serious risks for personal privacy. Accordingly, it is paramount to protect any sensitive data against recovery or deduction by third-parties to avoid the compromise of individual privacy. The most common security preservation practice is the use of cryptographic techniques. However, cryptographic protocols do not provide perfect security as the inference of critical information from non-critical ones remains a possibility. Indeed, the discovery of vulnerabilities of simple crypto-systems like that of the Needham-Schroeder public key protocol \cite{lowe1995attack} proved that cryptography is not enough to guarantee the privacy of information. Furthermore, the various techniques available are computationally intensive. This is why they cannot be immediately adopted in IoT where the network nodes are powered by battery. To facilitate the adoption of IoT in health-care, we need formal (preferably automated) verification of security properties. Formal verification entails the use of mathematical techniques to ensure that the system's design conforms to the desired behavior. Information flow properties are the most formal security properties. In fact, various ones have been defined in the literature including non-interference \cite{DBLPM82}, intransitive non-interference \cite{nbha1} and others (e.g. secrecy, and anonymity). Interested in confidentiality properties, we consider opacity, a general information flow property, to analyze IoT privacy in a heart attack detection system. Opacity's main interest is to formulate the need to hide information from a passive observer. It was first introduced in \cite{Mazar} and was later generalized to transition systems~\cite{Bryans2008}. It has since, been studied several times allowing the formal verification of system models (usually given as non deterministic automata or labeled transition systems). Its wide study led to the birth of several definitions (variants) as well as verification and enforcement techniques. If classified according to the security policy, then we are dealing with simple, $K$-step, initial, infinite as well as strong and weak opacity alongside their extensions (e.g, $K$-step weak and $K$-step strong opacity). The efforts of these studies also made possible not only opacity verification, but also its assurance via supervision \cite{Dubrieil2009}, \cite{SabooriH12} or enforcement \cite{Falcone2013}. A key limitation of these studies is that they have been very theoretical in the way they have approached and applied opacity. 

In this paper, we wish to bridge the gap between the theory of opacity and its practical application through the synthesis of an opaque IoT-based heart attack detection system. Building on the SOG-based verification approach developed in \cite{bourouis2015checking}, the purpose is to verify opacity in three of its forms (simple, $K$-step weak opacity and $K$-step strong opacity) to detect security violations in our synthesized system. Then to contribute an algorithmic approach that enforces simple opacity by padding the system with minimal dummy behavior.

This paper is organized as follows: Section 2 establishes all necessary basic notions including the SOG structure and the opacity property. In Section 3, we detail the case study. In Section 4, we illustrate the practical usefulness of the opacity verification approach in the heart attack detection system. Section 5 details our proposed approach to enforce simple opacity. Finally, we conclude in Section 6, and list some potential future works. 

%% file: preli.tex
\addcontentsline{toc}{section}{Preliminaries}
\section{Preliminaries}
Throughout this paper, the following notions are used to introduce our proposed approach.
Let $\Sigma$ be a finite alphabet.
\small
\begin{compactitem}
	\item $\Sigma^*$: is the set of all finite words over $\Sigma$, comprising the empty word $\varepsilon$;
	\item $L \subseteq \Sigma^*$: $L$ the language defined on $\Sigma^*$.
\end{compactitem}
The sequence $u$ is a prefix of  $v \in \Sigma^*$, denoted by $u \prec v$, if $\exists w$ s.t $u \cdot w = v$. We note $v-u=w$.
\subsection{Petri nets, WF-net and oWF-nets}
To model the services under consideration in our case study, we use Petri nets. A service can be considered as a control structure describing its behavior in order to reach a final state. We can represent it using a Workflow net, a subclass of Petri nets. A WF-net satisfies two requirements: it has one input place $i$ and one output place $o$, and every transition $t$ or place $p$ should be located on a path from place $i$ to place $o$. To model the communication aspect of a service, we can use another variant of Petri nets called open Work-Flow nets which is enriched with communication places representing the (asynchronous) interface. Each communication place represents a channel to send or receive messages to or from another oWF-net. 
\begin{definition}[oWF-net \cite{Massuthe05anoperating}] \mbox{} \\
\small
	An open Work-Flow net is defined by a tuple $\mathcal{N} = (P, T, F,W,m_0, I,O,m_f)$:
	\vspace{-0.3cm}
	\begin{itemize}
		\item $(P, T, F, W)$ is a WF-net;
			\begin{compactitem}
			\item $P$ is a finite set of places and $T$ a finite set of transitions;
			\item $F$ is a flow relation $F\subseteq  (P\times T) \cup  (T\times P);$
			\item $W: F \rightarrow \mathbb{N}$ is a mapping allocating a weight to each arc.
			\end{compactitem}
		\item $m_0$ is the initial marking;
		\item $I$ is a set of input places and $O$ is a set of output places ($I\cup O$: the set of interface places). 
		\item $m_f$ is a final marking.
	\end{itemize}
\end{definition}
\vspace{-0.3cm}
Having the same semantics as Petri nets, the behavior of WF-nets and oWF-nets can be represented by Labeled Transition Systems (LTS), a more general model than the reachability graph.

\subsection{Labeled Transition System}
An LTS is defined as follows:
\begin{definition}[Labeled Transition System] \mbox{} \\
\small
A Labeled Transition System is a  $4$-tuple $\mathcal{G}=(Q,q_{init},\Sigma,\delta)$:
	\small
	\vspace{-0.3cm}
	\begin{itemize}
		\item $Q:$ a finite set of states;
		\item $q_{init}:$ the initial state;
		\item $\Sigma:$ actions' alphabet;
		\item $\delta: Q \times \Sigma  \rightarrow Q:$ the transition function where: $q,\; q' \in Q$ and $\sigma \in \Sigma$, $\delta (q,\sigma)=q'$ meaning that an event $\sigma$ can be executed at state $q$ leading to state $q'$.
	\end{itemize}
\end{definition}
\vspace{-0.3cm}
The language of an LTS $\mathcal{G}$ is defined by $L(\mathcal{G})=\{t \in \Sigma^*, q_0 \xrightarrow{t} q_f\}$. An LTS can be considered as an automaton where all states are accepting final states. Therefore, the language accepted by the LTS is prefix-closed.

To reflect the observable behavior of an LTS, we specify a subset of events $\Sigma_o \subseteq \Sigma$ and $\Sigma - \Sigma_o = \Sigma_{u}$ where $\Sigma_o$ is the set of events visible to a given observer and $\Sigma_{u}$ is the set of events which are invisible to said-observer. The behavior visible by an observer is defined by the projection $P_{\Sigma_o}$ from $\Sigma^*$ to $\Sigma_o^*$ that removes from a sequence in $\Sigma^*$ all events not in $\Sigma_o$. Formally, $P_o$: $\Sigma^*$ $\rightarrow \Sigma^*_o$ is defined s.t.:\\
$ \begin{cases}
P_{\Sigma_o}(\epsilon)=\epsilon; \\
P_{\Sigma_o}(u \cdot \sigma)= \begin{cases}
P_{\Sigma_o}(u) \:\text{if}\: \sigma \notin \Sigma_o;\\
P_{\Sigma_o}(u)\cdot \sigma \;\text{otherwise}.
\end{cases} $ 
 \text{Where:} $\sigma \in \Sigma \: \text{and} \: u \in \Sigma^*$.
$ \end{cases}$
\subsection{Opacity}
Opacity's main interest is in capturing the possibility of using observations and prior-knowledge of a system's structure to infer secret information. It reflects a wide range of security properties. Opacity's parameters are a secret predicate, given as a subset of sets or traces of the system's model, and an observation function. This latter captures an intruder's abilities to collect information about the system. A system is, thus, opaque w.r.t. the secret and the observation function, if and only if for every run that belongs to the secret, there exists another run with a similar projection from the observer's point of view and that does not belong to the secret \cite{Dubrieil2009, falcone, FengLin}.
In this paper, we focus on 3 opacity variants as defined by the authors in \cite{falcone}: simple, $K$-step weak and $K$-step strong opacity.
\begin{definition}[Simple opacity \cite{falcone}] \mbox{} \\
\small
	Given an LTS $\mathcal{G}=(Q,q_{0},\Sigma,\delta)$ with $\Sigma_{o} \subseteq \Sigma$ is the set of observable events and $S\subseteq Q$ is the set of secret states.
	The secret $S \subseteq Q$ is opaque under the projection map $P_{\Sigma_{o}}$ ou $(G, P_{\Sigma_{o}})-opaque$ iff: $ \forall u \in L_{S}(G), \exists v\in L(\mathcal{G}): (v\;\approx_{\Sigma_{o}} \: u) \wedge (v \notin L_{S}(G)).$
	\label{def:simpleop}
\end{definition}

While simple opacity deals with the non-discloser of the fact that the system is currently in a secret state, $K$-step opacity deals with the non-discloser of the fact that the system was in a secret state in the past. $K$-step weak opacity ensures that the system wasn't in a secret state $K$ observable events ago, while $K$-step strong opacity formulates the need to make sure that, $K$-steps backwards, the system does not end, and have not crossed any secret states.

\subsection{Symbolic Observation Graph}
The SOG is an abstraction of the reachability graph. It is constructed by exploring a system's observable actions which are used to label its edges. The unobservable actions are hidden within the SOG nodes named aggregates. Binary Decision Diagrams (BDDs) \cite{bryant92} are used to represent and efficiently manage the SOG nodes.
The definition of an aggregate and that of the SOG are given in the following:
\begin{definition}[Aggregate] \mbox{} \\
\small
	Given an LTS $\mathcal{G}\:=(Q,q_{0},\Sigma,\rightarrow, \delta)$ with $\Sigma = \Sigma_{o} \cup \Sigma_{u}$. An aggregate $a’$ is a non empty set of states satisfying: $q\in a \Leftrightarrow Saturate(q)\subseteq a$ where: $Saturate(q)=\{ q' \in Q :  q \xrightarrow{w} q' \text{and} \: w \in \Sigma^{*}_{u}\}.$
\end{definition}

\begin{definition}[Deterministic SOG] \mbox{} \\
\small
A deterministic $SOG (\mathcal{A})$  associated with an LTS $\mathcal{G}= (Q,q_{0},\Sigma_{o}\cup\Sigma_{u},\delta)$ is an LTS  $(A,a_{0},\Sigma_{o},\Delta)$ where:
\vspace{-0.3cm}
\begin{enumerate}
\item \label{def:SOG:dnodes}$A$ a finite set of aggregates with:
\begin{enumerate}
\item \label{def:SOG:dnodes1} $a_0 \in A$ is the initial aggregate s.t. $a_0=Saturate(q_{0}) \text{;}$
\item \label{def:SOG:dnodes2} For each $a \in  A$, and for each $\sigma\in \Sigma_{o}$, $\exists q\in a, q'\in Q \colon q
\xrightarrow{\sigma} q' \Leftrightarrow \exists a'\in A:\; a'=Saturate (\{q'\in Q, \exists q\in a \: with \:
q \xrightarrow{\sigma}q'\}) \wedge (a,\sigma,a') \in \Delta$;
\end{enumerate}
%\item \label{def:SOG:dlabels} $\Sigma_{A}=\Sigma_{o}$;
\item \label{def:SOG:darcs} $\Delta \subseteq A\times
\Sigma_{o} \times A$  is the transition relation.
\end{enumerate}
\end{definition}
\vspace{-0.3cm}
The SOG allows the on-the-fly-verification of opacity variants: simple, $K$-step weak and $K$-step strong opacity \cite{bourouis2015checking}. It also allows the detection of cases of opacity violation.  

%% file: case.tex
\addcontentsline{toc}{section}{Motivating Scenario}
\section{Motivating Scenario}
Heart disease is the first cause of morbidity and mortality in the world, accounting for 28.30\% of total deaths each year in Tunisia alone \cite{Organization2014}. Investment in preventive health care such as the use of IoT monitoring devices and tools may help lower the cost of processing and the development of serious health problems. In fact, integrating clinical decisions with electronic medical records could decrease medical errors, reduce undesirable variations in practice, and improve patient outcomes.

Our case study considers IoT integration with cloud computing. We use a connected bracelet, fog nodes, a private and a public Cloud, and a mobile application, which together form a medical application. This latter provides continuous monitoring of the vital data of a given patient. Regular or routine measurements could help to detect the first symptoms of heart malfunction, and makes it possible to immediately trigger an alert. The vital information collected by the bracelet worn by the patient includes cardiac activity, blood pressure, oxygen levels and, temperature. As mentioned earlier, we consider an IoT application in a hybrid cloud/fog environment. The cloud \cite{zhang2010cloud} is considered as a highly promising approach to deliver services to users, and provide applications with low-cost elastic resources. Given the fact that IoT  suffers limited computational power, storage capacity and bandwidth, cloud computing ease the issues in enabling the access, the storage, and the processing of the large amount of generated data.

Public clouds provide cheap scalable resources. Making it useful for analyzing the patient's vital data which would be costly as it requires extensive computing and storage resources. However, we must take into account that storage of health records on a public environment is a privacy risk. To avoid such security leaks, we could deploy the application on a secure private cloud. But seeing this latter's limited resources, this may degrade the overall performance. To prevent this, the workflow can be partitioned between a private cloud and a public one. Therefore, the confidential medical data will be processed on the private cloud. Other workflow actions can be deployed on the public cloud dealing with anonymized data.

The use of a cloud-based framework poses the problem of delay when sending and receiving data between the objects and geographically far cloud resources thus jeopardizing the patients' well-being given that triggering timely responses is the purpose of this data. To resolve this issue, data gathering can be moved from the cloud domain to that of the fog \cite{BonomiMZA12}. Bringing this action closer to the connected object shortens the transmission time, and reduces the amount of data transferred to the cloud.
The proposed workflow is described as follows:
	\begin{compactitem}
		\item A patient may register via the mobile app by entering his information. This information include personal data and medical history (personal and family medical histories, surgical history, drug prescriptions, and the doctors' notes). 
		\item The patient's medical history is then transmitted to the private cloud. After reception, this latter anonymizes the data by stripping off all that could identify the patient leaving only medical data, which it sends to the public cloud.
		\item The public cloud receives the anonymized data, and proceeds to the classification attaching to each medical file a class.
		\item The patient is equipped with a  measuring bracelet connected to the processing components (Fog nodes). The data sent to the fog domain is a set of vital data recorded over a period of time. 
		\item The fog node collects the data then compares it to its predecessors, searching for any vital signs changes. When the node determines that a change has occurred, it sends the data to the private cloud. 
		\item The private cloud links the gathered data with the patient, transmitting this data and the class ascribed to the patient, to the public cloud.
		\item The public cloud reads the data, analyzes it, and then provides results. When the risk of heart attack is detected, it immediately notifies the patient's app.
	\end{compactitem}

%% file: verif.tex
\addcontentsline{toc}{section}{Modeling and Verification}
\section{Modeling and Verification}
The case study contains five services, namely, a connected bracelet (Br), a fog node (Fog), a private cloud (CPr), a public cloud (CPub), and a smartphone application (App). Figure \ref{fig:owfnet} depicts the oWF-nets of the Br, Fog, CPr, CPub and the App, respectively. We note that the transitions entailing the sending (respectively reception) of a messages are indicated by adding a ! (respectively a ?) mark.
\input{owfnet.tex}
In this case study we want to illustrate the ability of the SOG-based verification approach to meet privacy demands. The first step is to create the underlying LTS of each oWf-net. Secondly, we identify the observable and unobservable actions of each net as well as the secret states. Then, we build the SOG models from each net’s LTS using the data from the previous step verifying, at the same time, their opacity.

The Br workflow (Figure \ref{fig:brnet}) starts by collecting data ($T_1$) through the sensors mounted in the bracelet, which will then be sent to the closest Fog node. Next it creates the message comprising the data ($T_2$) and sends this message ($T_3?$). Not having any security requirements for the bracelet, thus, there is no need to check its opacity.

The Fog WS (Figure \ref{fig:fognet}) has an internal set of operations, and a set of external cooperative ones. The second set concerns the exchanges of the fog with the Br, the CPr and the App. After receiving the data ($T_1!$), we consider two scenarios. The first is when the Fog communicates for the 1st time with the bracelet ($T_3$). In this case, it sends a request ($T_5?$) to the App to retrieve data from the patient's medical history. Then, it will receive these data through ($T_6!$). The second scenario begins by selecting the last recorded data ($T_4$). The next step is to compare ($T_7$) the data retrieved by one of the mentioned scenarios with the data sent by the Br. When the node detects a change in values ($T_9$), it will immediately transmit the data to CPr ($T_{10}?$). If there is no change ($T_8$), the Fog doesn't perform any processing. Finally, the new data will be stored locally in the Fog ($T_{11}$). To ensure the
privacy of fog secret information, we define the secret state $S=\{S_6\}$ which is related to receiving patient's medical history. To conform with the security needs, the observable transitions of the Fog are $\Sigma_{o} = \{T_1!,T_5?,T_6!,T_{10}?\}$, while
the unobservable part is $\Sigma_{u}=\{T_2,T_3,T_4,T_7,T_8,T_9,T_{11}\}$. Using this data, and after creating the LTS, we proceed to the opacity verification which is done on-the-ﬂy, while creating the SOG-abstraction of the model. We get the SOG in Figure \ref{fig:sogfog} and we note that we have a single secret state belonging to an aggregate which holds other, non secret, states. We can conclude, then that the fog's SOG is both simple, and $K$-step weakly and strongly opaque.

The CPr worflow (Figure \ref{fig:cprnet}) contains two scenarios. The first one starts by receiving the data of a new registered patient ($T_1!$). The CPr subsequently proceeds with the recording ($T_2$) and the anonymization ($T_3$) of the received data. The anonymised data will then be transmitted to the CPub ($T_4?$). After receiving ($T_5!$) the class, this latter is associated with the patient ($T_6$). The second scenario starts when the CPr receives ($T_7!$) the data sent by the Fog. The CPr combines the data with the patient by searching for its unique identifier (ID)($T_8$). If the ID cannot be found ($T_9$), the CPr sends a request to the App so that the patient re-enter his personal information ($T_{10}?$). Thereafter, it receives the requested data ($T_{11}!$) and it  pursues the first scenario. For the second case, when the ID is found, the CPr transmits the data and the class to which the patient belongs to the CPub ($T_{13}?$). Afterwards, the CPr receives and records respectively 3 types of messages, each one belongs to an alert type: low ($T_{14}!$ \& T15), medium ($T_{16}!$ \& $T_{17}$) and high ($T_{19}!$ \& $T_{20}$). To protect the privacy of individual patients, the CPr need to hide the update procedure performed on the patient's personal information. It must keep secret the states related to the new patient registration ($S_4$ \& $S_{16}$) and the anonymization of his data ($S_7$ \& $S_{21}$). It is also required to withhold secret the states related to sending alerts when the symptoms of a heart attack are detected ($S_{22}$ \& $S_{23}$). So the set of secret states for the CPr is $S=\{S_1,S_3,S_4,S_7,S_16,S_{21},S_{22},S_{23}\}$, where $S_1$ stands for the marking related to the reception of the data sent by the fog, while $S_3$ reflects that related to patient ID search. To conform with these needs, the observable transitions of the CPr are $\Sigma_{o}=\{T_1!,T_4?,T_5!,T_7!,T_{10}?,T_{11}!,T_{13}?,T_{14}!,T_{16}!,T_{18}?,T_{19}!,T_{21}?\}$, while the unobservable ones are $\Sigma_{u}=\{T_2,T_3,T_6,T_8,T_9,T_{12},T_{15},T_{17},T_{20}\}$. With this configuration, we conduct the opacity verification and get the SOG in Figure \ref{fig:sogcpr}. Thus, the CPr workflow is not opaque and is not k-step weakly and strongly opaque. Indeed, the two secret states $S_{22}$ and $S_{23}$, each belonging to an aggregate that doesn't hold other non-secret states. An attacker can then disclose secret information after the traces $T_7T_{13}T_{16}T_{18}$ and $T_7T_{13}T_{19}T_{21}$. The CPr service is therefore unsafe and needs to be improved to adapt it to handle private data.
\input{sog.tex}
Taking into account that the public cloud is available for public use, we don't have secrets to be hidden from an external observer. So, in the following, we will only describe the CPub actions (Figure \ref{fig:cpubnet}) and we won't proceed the opacity verification. The first set of CPub operations concerns the internal operations which include the processing of the data sent by the CPr: the classification ($T_2$) and the prediction ($T_5$) which in our case aims to detect the risk of heart attack. As regards the external operations, the CPub receives two messages from the CPr. The first one ($T_1$) includes the anonymised data and the second ($T_4$) includes the data collected by the Br and the class to which the patient belongs. In response to the received messages, the CPub sends the classification result to the CPr ($T_3$) and sends 3 types of alerts according to the results obtained from the prediction: $T_6$ for the low alert, ($T_7$ for the medium alert and $T_8$ for the high alert.

The last service is that of the App (Figure \ref{fig:appnet}). The set of internal operations of the App consists of two actions. The first action concerns the notification ($T_9$) through which the application warns the patient. The second concerns the application to register ($T_1$) which allows a new patient to deposit his information. After registration, the provided information will be sent ($T_2$) to the CPr. The App shares patient information with the Fog ($T_3!$ \& $T_4?$) when this latter communicates for the first time with the Br. It also shares the medical history with the CPr ($T_5!$ \& $T_6$) when it fails to find the patient ID. At the end, the App receives two types of alerts ($T_7?$ for the medium alert and $T_8?$ for the high alert) when the risk of a heart attack is detected. The App must be opaque to both sides of the communication with regards to its set of secret states when dealing with either the CPr or the Fog. To match these needs the
observable transitions of the App is  $\Sigma_{o}=\{T_2?,T_3!,T_4?,T_5!,T_6?,T_7!,T_8!\}$, while the unobservable ones are $\Sigma_{u}=\{T_1,T_9\}$. Moreover, the set of secret states are $S=\{S_2,S_6,S_7,S_9,S_{10},S_{11}\}$, with $S_2$ is the marking related to the request to register of a new patient, $S_6$ is that related to the sending of patient data, $S_7$ is that triggered due to the sending of personal information of a new patient, $S_{11}$ is related to the sending of medical history, and finally $S_9$ and $S_{10}$ reflect the secrets associated with sending the notification to warn the patient. Conducting the opacity verification for the App, we obtain the SOG depicted in Figure \ref{fig:sogapp}. We say that the App SOG is not opaque according to Definition \ref{def:simpleop}. Consequently it is not
$K$-step weakly, and strongly opaque.

%% file: owfnet.tex
\begin{figure*}[!htbp]
	\begin{minipage}{0.2\textwidth}
		\begin{subfigure}[Bracelet oWF-net.]{
				\centering
				\includegraphics[width=0.9in]{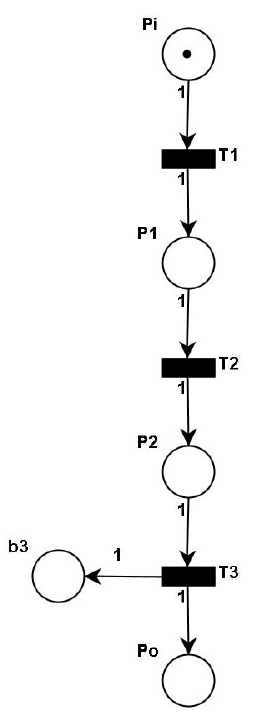}
				\label{fig:brnet}
			}
		\end{subfigure}
		\vfill
		\begin{subfigure}[Fog oWF-net.]{
				\centering
				\includegraphics[width=1.8in]{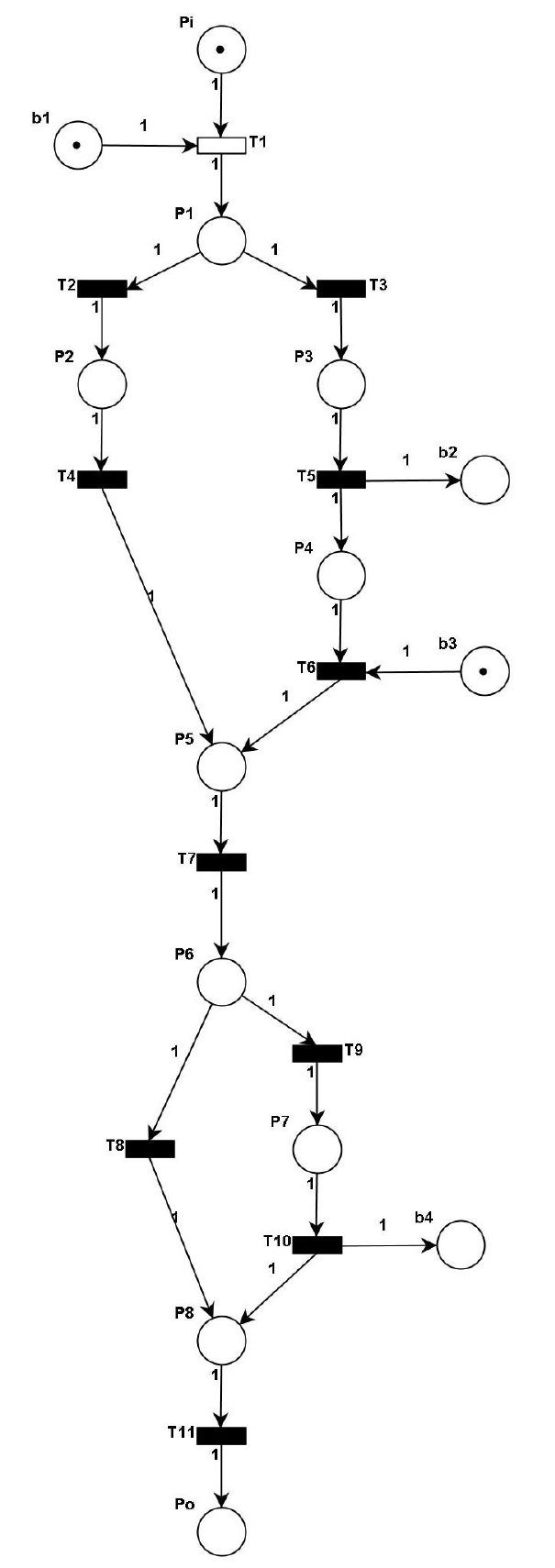}
				\label{fig:fognet}	
		}
		\end{subfigure}
	
	\end{minipage}
	\hspace{2.7cm}
	\begin{minipage}{0.2\textwidth}
			
		\begin{subfigure}[Private cloud oWF-net.]{
				\centering
				\includegraphics[width=2.9in]{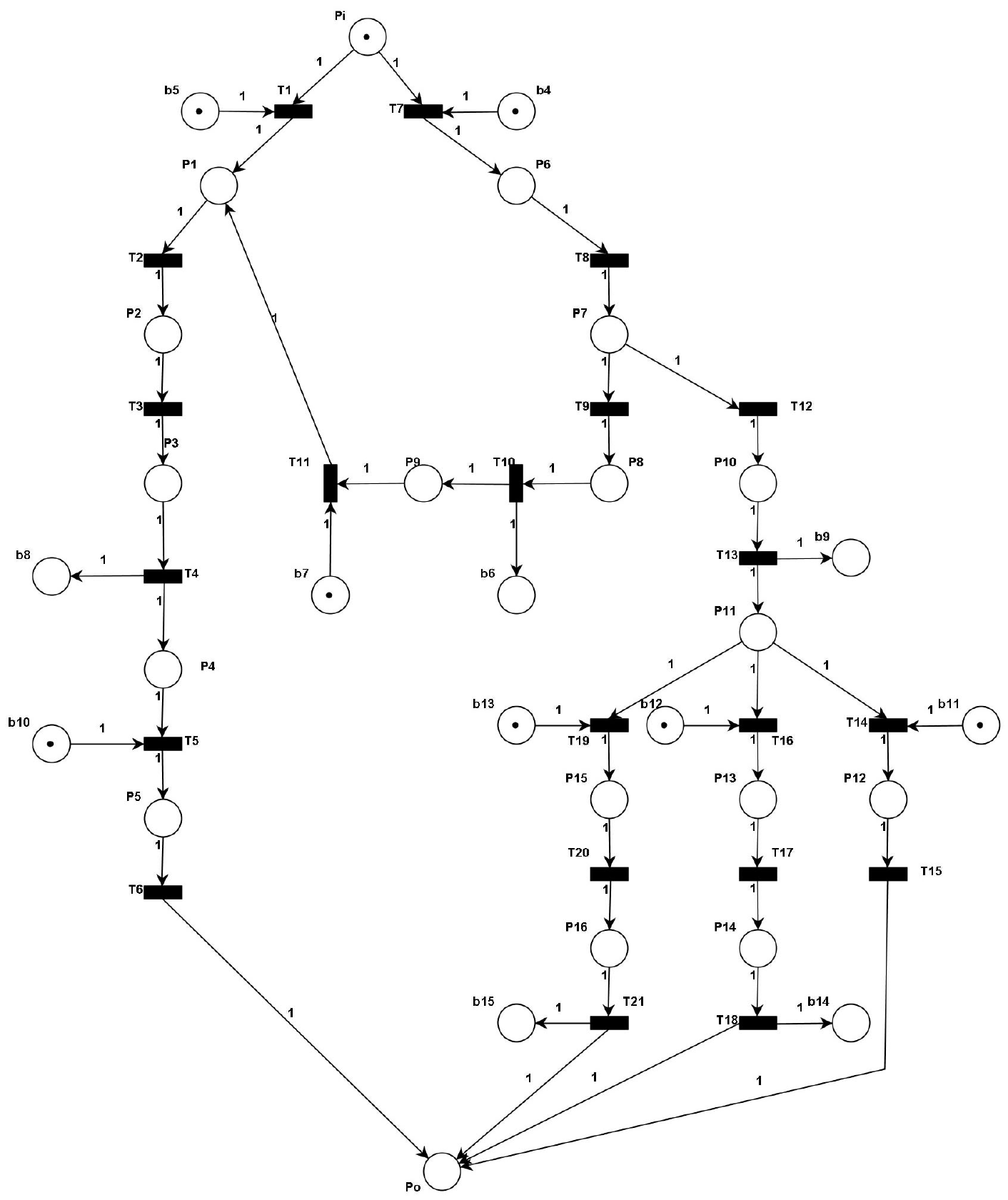}
				\label{fig:cprnet}
		}
		\end{subfigure}
	\vfill
		\begin{subfigure}[Public cloud oWF-net.]{
				\centering
				\includegraphics[width=2.9in]{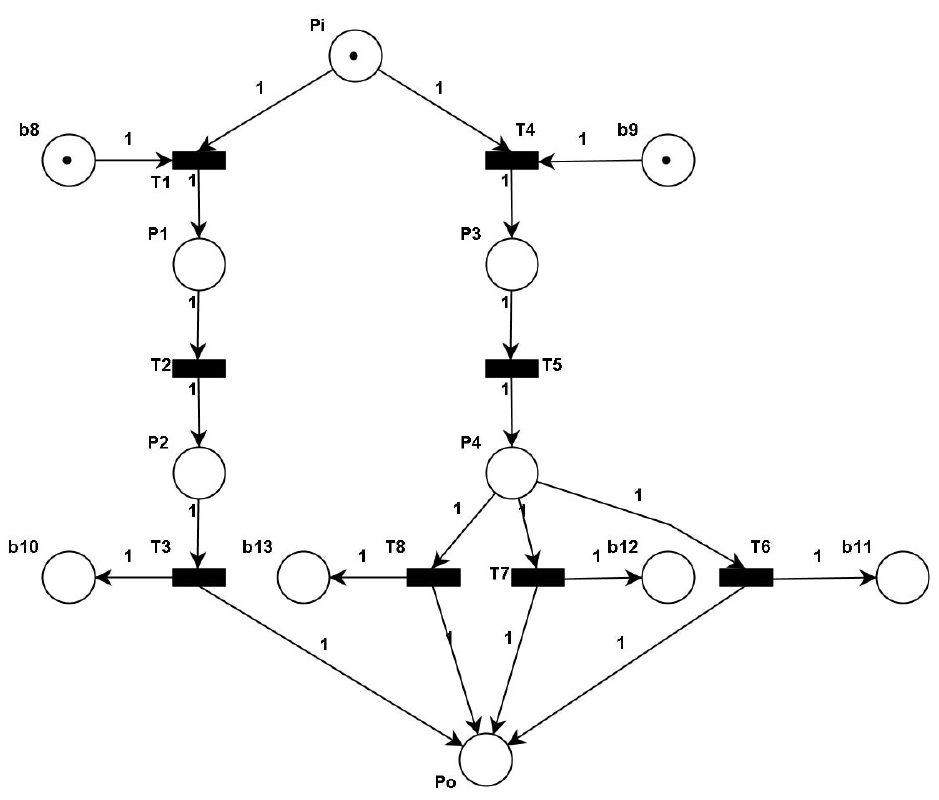}
				\label{fig:cpubnet}
			}
		\end{subfigure}
		\vfill
		\begin{subfigure}[Application oWF-net.S]{
				\centering
				\includegraphics[width=2.5in]{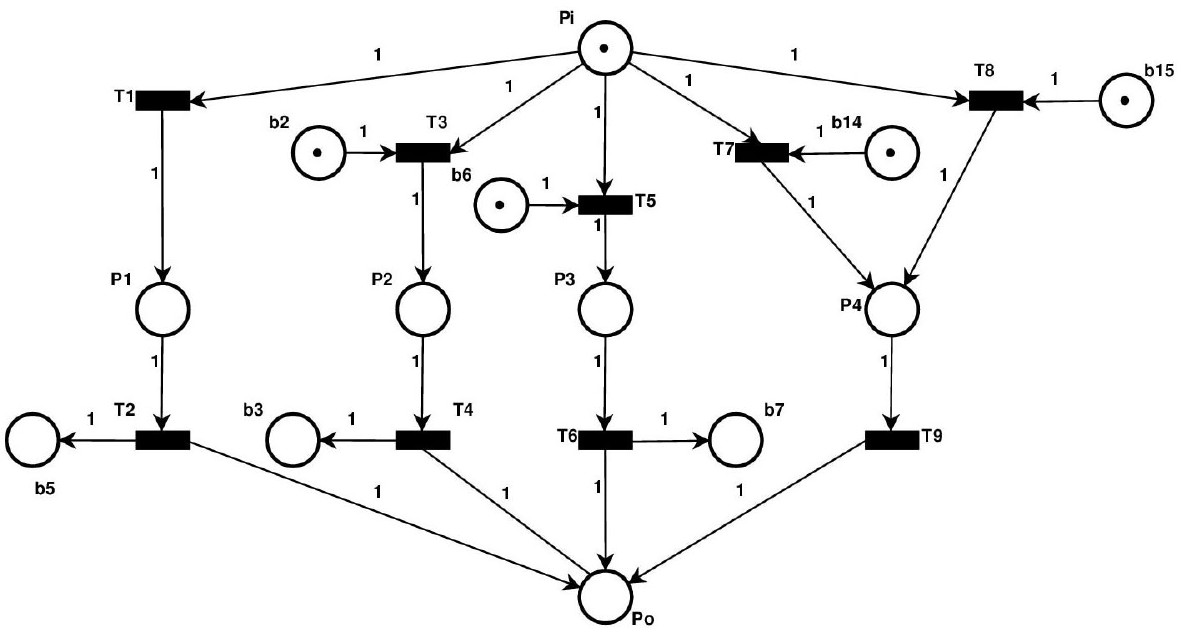}
				\label{fig:appnet}
			}
		\end{subfigure}
	\end{minipage}
	\caption{Case Study oWF-nets.}
	\label{fig:owfnet}
\end{figure*}

%% file: sog.tex
\begin{figure*}[!htbp]
\begin{minipage}{0.2\textwidth}
	 \begin{subfigure}[ The SOG of the Fog Node.]{
			\centering
			\includegraphics[width=2.5in]{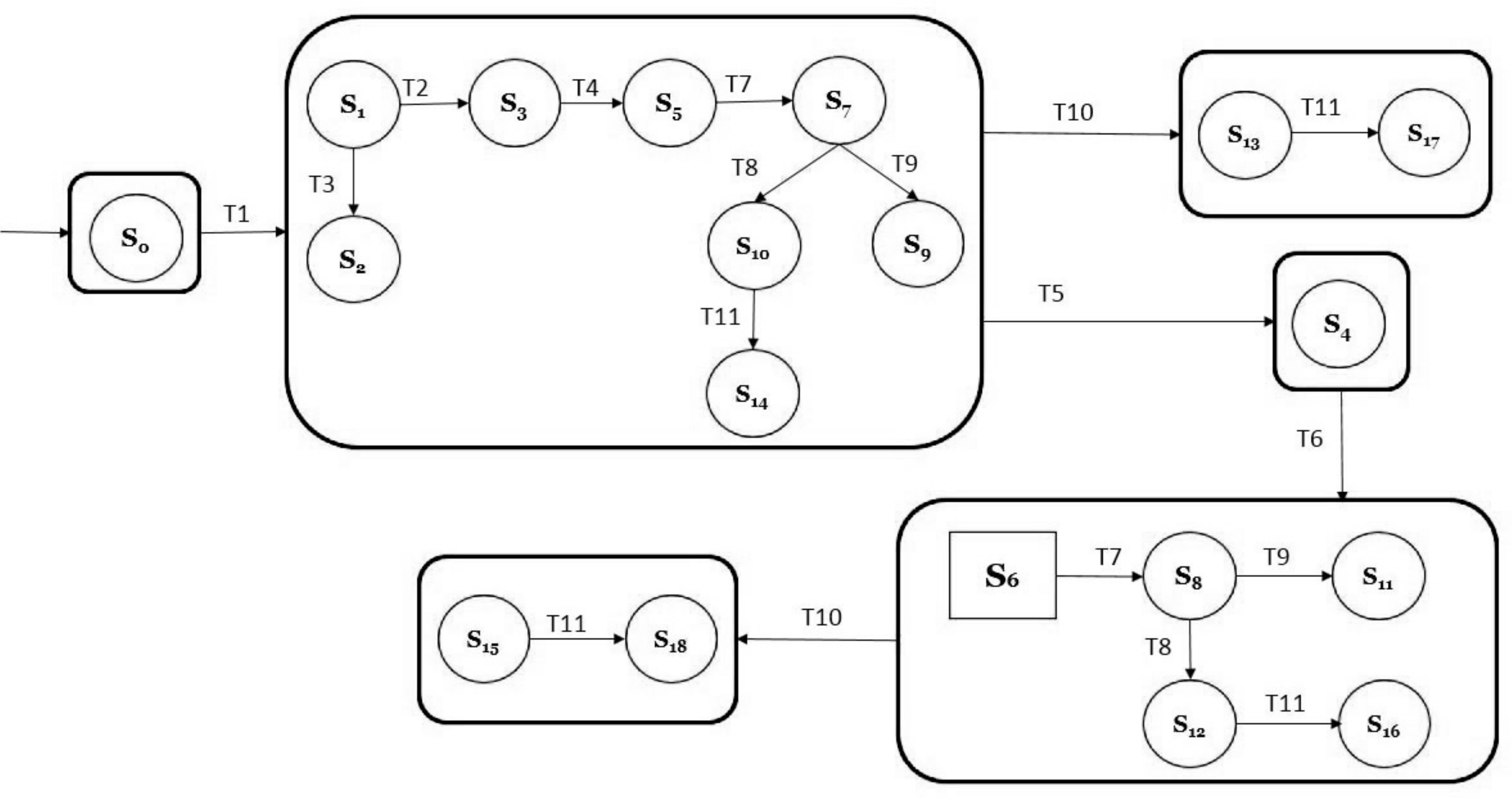}
			\label{fig:sogfog}
		}
	\end{subfigure}
	\hfill
	\begin{subfigure}[ The SOG of the Private Cloud.]{
			\centering
			\includegraphics[width=2.5in]{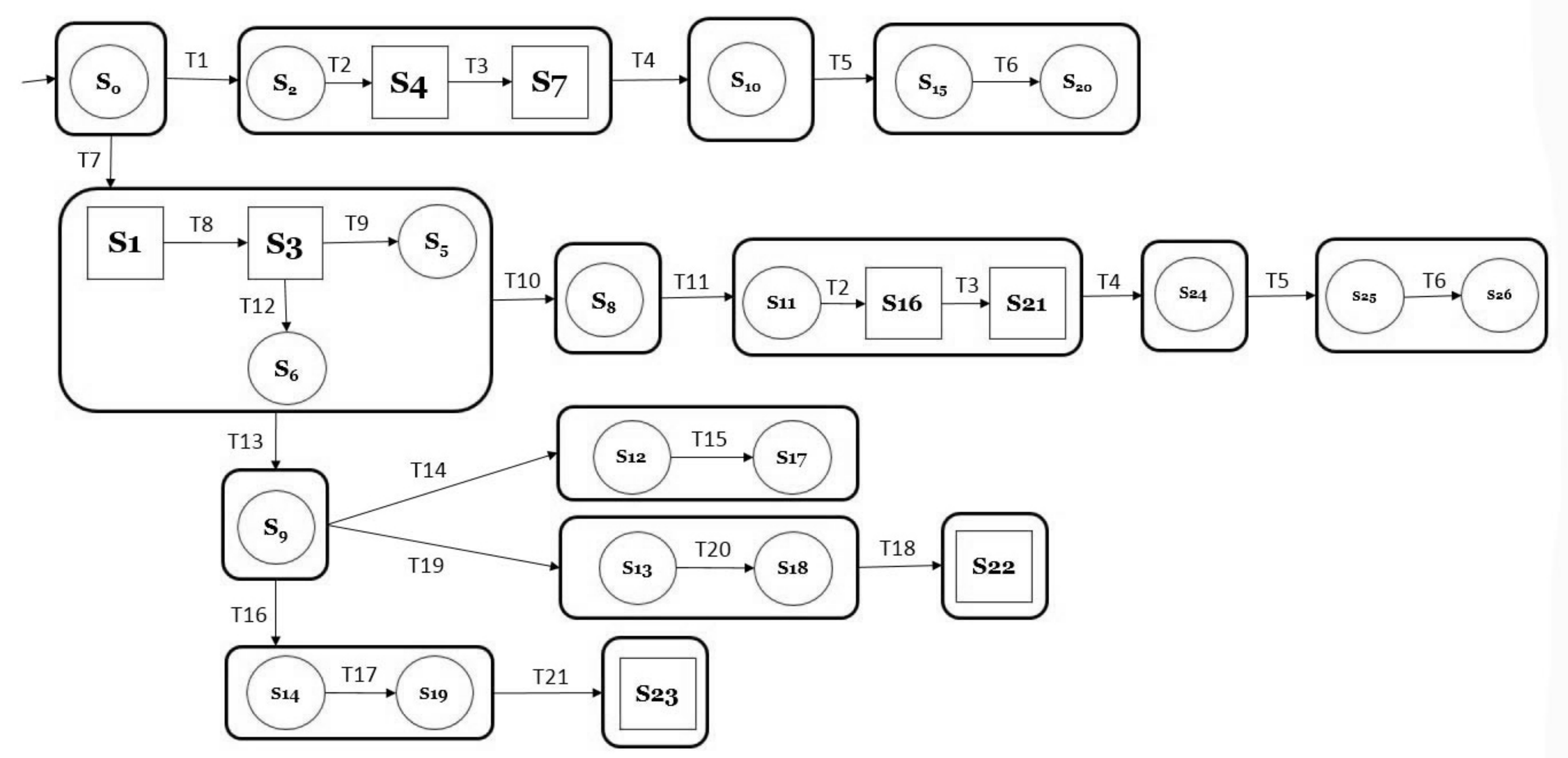}
			\label{fig:sogcpr}
		}
	\end{subfigure}
\end{minipage}
\hspace{4.2cm}
\begin{minipage}{0.2\textwidth}
\begin{subfigure}[ The SOG of the Mobile Application.]{
		\centering
		\includegraphics[width=2in]{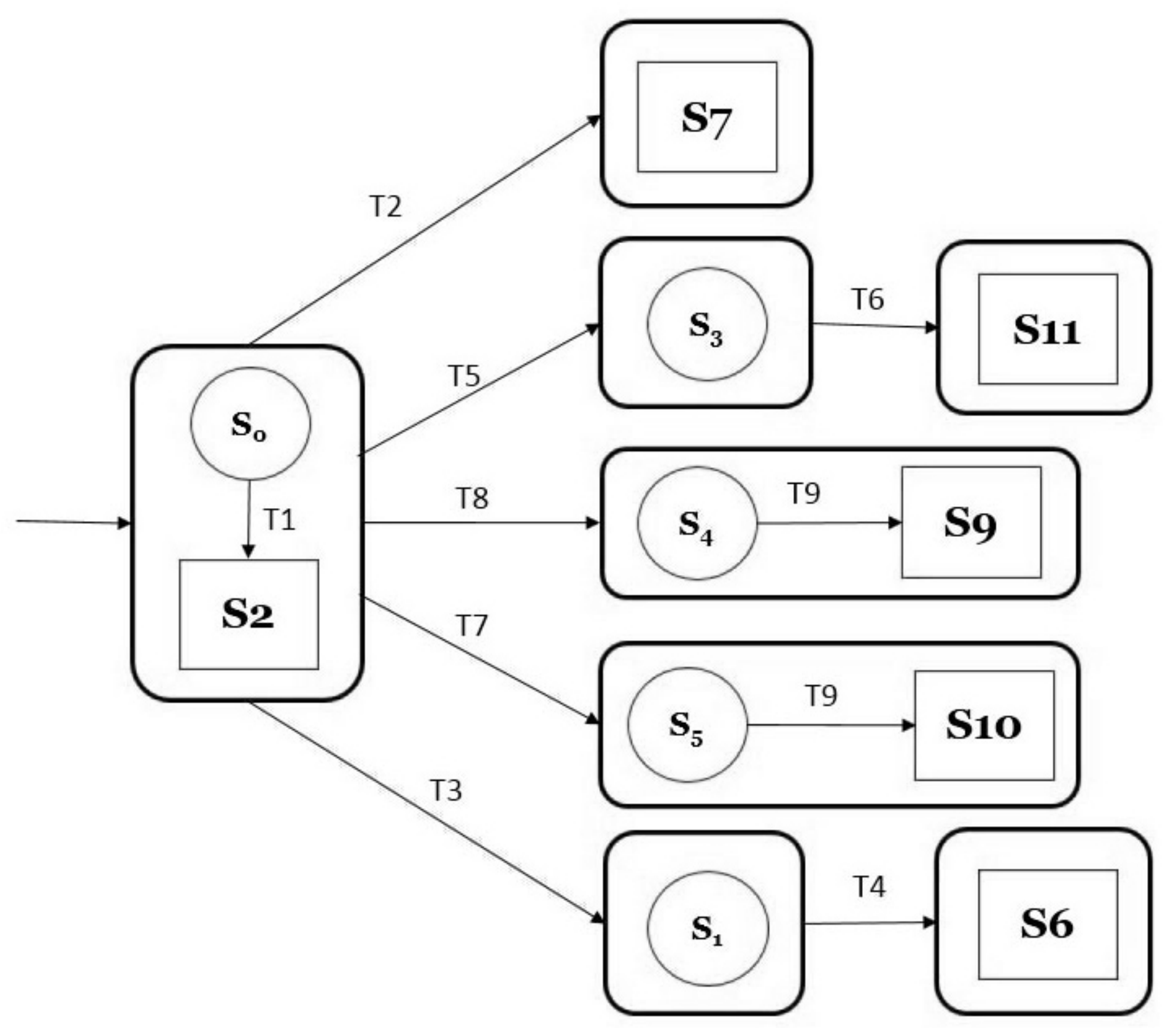}
		\label{fig:sogapp}
	}
\end{subfigure}
\end{minipage}
	\caption{The SOGs of the Case Study WSs.}
\label{fig:sog}
\end{figure*}

%% file: app.tex
\addcontentsline{toc}{section}{SOG-based Enforcement of Opacity}
\section{SOG-based Enforcement of Opacity}
In this section, we describe the opacity enforcement problem introducing algorithms to secure the heart attack detection system. Considering a language $L$ and a secret language $L(\varphi) \in L$, when opacity fails of a secret $\varphi$ for a finite system $S$, we provide an effective method to synthesize automatically a system $S'$ obtained by minimally modifying the system S so that the secret $\varphi$  is opaque for $S'$. To synthesize $S'$, we focus on language modification. If a secret language $L(\varphi)$ is not opaque for a system behavior described by the language $L(S)$, we can modify the behavior by padding it with dummy behaviors. We can then extend the language by computing a minimal super-language of $L$. In \cite{Yeddes16}, the author has derived an algorithm to compute $min \; \prod_{super}^\varphi$ to assist the designer develop a system that satisfies the opacity property for a secret language.
\begin{theorem} \cite{Yeddes16}
	\label{theorem:minimal}
	Let a language $L$ defined on an alphabet $\Sigma = \Sigma_{o} \cup \Sigma_{u} $ and a static projection $\pi_O$ defined above on the same alphabet and a secret $\varphi \subseteq L$, then:
	\centering $min\; \prod_{super}^\varphi(L) = L \cup (\pi_o(\varphi) \backslash (\pi_o(\varphi)\cap \pi_o(L \backslash \varphi)))$	
\end{theorem}

The proposed approach builds upon the SOG structure to check the system's opacity. If the system is not opaque, the SOG construction allows for detection of all opacity violations provided as a counterexample. These counterexamples will later be used to improve the system security (opacity) by locating the paths leading to the disclosure of private information and performing necessary changes that would render it opaque. 
Then we compute the minimal super-language that provides us with the restricted language to be added in order to modify the system behavior. For each incident of opacity violation, we match a trace among the calculated super-language and an unobservable event will be added to this trace. In order to opacify the system, we apply the backtracking method. We implement adjustments where needed to the SOG and the LTS and we thus return to the starting model, the Petri net. 

\subsection{The SOG-based Algorithm for the Verification of Simple Opacity}
The use of SOG-based algorithm in the verification of simple opacity proved efficient \cite{bourouis2015checking}. This is due to the symbolic representation of the aggregates (employing BDDs), and to the on-the-fly verification (Opacity is verified while constructing the SOG). The SOG construction is stopped when the property is proven unsatisfied and a trace (counterexample) that violates the opacity is supplied. To adopt this algorithm for our enforcement approach, we will bring necessary modifications to it.
\input{sogalgorithm.tex}

The verification is performed on LTS-modeled systems. Taking into account that we are trying to opacify Petri nets, the first modification needed to the algorithm presented in \cite{bourouis2015checking} consists in replacing the input by a Petri net-modeled system. The petri net has 2 sets of transitions: observable and unobservable actions, and a set of secret marking subsequently representing the states judged to be secret in the LTS.  We add in line 3 a Stack, namely $CounterExample$  with all the standard functions ($push$, $pop$ and $top$), whose elements are quadruples composed by the counter-examples, a transition $t$, an actual aggregate $a$ and an aggregate $a'$, successor of $a$ by $t$. Then, the algorithm 1 starts by constructing (line 6) the reachability graph which represents the LTS. Once other changes have been made (i.e. line 10 \& 29), when the opacity is violated, neither the verification nor the construction of the SOG stops. All the paths leading to the disclosure of privacy are stacked into $CounterExample$. Once all nodes are explored and the SOG construction is finished, and if the stack is not empty we proceed to opacification.
\subsection{The Opacification Proposed Algorithm}
The opacification algorithm has a pretty straightforward mechanism. It begins by computing the minimal super-language ( Theorem \ref{theorem:minimal}). The next step consists in recuperating (line 4) the first elements of the stack ($CounterExample$). Next, the algorithm goes through the $foreach$ loop which takes each word of the calculated super-language. If such a word is equivalent with the trace recuperated from the stack, then we proceed to opacify the SOG. We begin by creating (line 8) a new state $q_{new}$ that we will add (line 9) into the aggregate $a'$.
At line 11, we pass to opacify the LTS. We retrieve the last state $q$ included in the aggregate $a'$. A new unobservable transition $t_{new}$ will be created. Then, the algorithm inserts (line 13) the new state $q_{new}$ to the LTS states, adds (line 14) the new transition $t_{new}$ to the set of unobservable events $\Sigma_{u}$, and defines the transition function between $q$, $t_{new}$ and $q_{new}$. 
Starting from line 16, the algorithm performs the Petri net opacification by creating, at first a new place $p_{new}$ and adding it to the set of places. It also adds the transition $t_{new}$ to the set of transitions. To specify the flow relation between $p$, $t_{new}$ and $p_{new}$, the algorithm adds an arc for each relation and assigns to each arc a weight. Afterwards, it modifies the incidence matrix. Finally, the algorithm pops the stack and restarts the operations until the final emptying of the stack presenting the ending test of the while loop.

Being a particular type of Petri nets, oWF-nets require different method of opacification. When fetching the place $p$ (the execution of $getPlace$), we have to exclude the output places. Furthermore, oWF-nets require only one final place $p_o$. So, following the addition of the unobservable transition $t_{new}$, we must escape adding the new place. And a flow relation will be added between $t_{new}$ and $p_o$. Other specific case that may be necessary,  when the place returned by $getPlace$ is a destination place, we require further changes on the oWF-net. The first step is to retrieve the transition that following its crossing marked the output place. Step two is to delete the flow relation between $t$ and $p_o$. The following step is to create a new place $p_{new}$ and to add the unobservable transition $t_{new}$. Then, we create the flow relations between $t$, $p_{new}$, $t_{new}$ and $p_o$.
\subsection{Case Study Opacification}
Through the case study presented earlier, we will show the practical use of our approach. In this section, we will describe the changes we are making to render the CPr workflow opaque.

Conducting the opacity verification, we noted that the CPr workflow is not opaque and the algorithm 1 returned two counterexamples: 
$CounterExample=\{
(T_7T_{13}T_{19}T_{18},$\\$ \; \{S_{13},S_{18}\}, \; T_{18}, \; \{S_{22}\}), \;
(T_7T_{13}T_{16}T_{21}, \; \{S_{14},S_{19}\}, \; T_{21}, \{S_{23}\}) \}$. \\
As the opacification algorithm starts by computing the minimal super-language, the function $ComputationMinSL$ returns the following result: \\ $minSl=\{T_7T_{13}T_{19}T_{18},\; T_7T_{13}T_{16}T_{21}\}$.

The next step of our opacification approach is to bring the necessary modifications to the SOG and then we return to the LTS and to the Petri-net. To treat the first opacity violation for the CPr, the algorithm 2 starts by creating a new state $q_{new}=S_{27}$. This is added to the aggregate that contains $S_{22}$. Then, a new unobservable transition is created ($t_new = T_ {22}$) and  the relation between $S_{22}$, $T_{22}$ and $S_{27}$ is defined. 
The final step is the opacification of the oWF-net. To do this, we applied the specific case of algorithm 2. We start first by removing the flow relation ($T_{21} × P_o$). Thereafter, we add the place $P_ {17}$. Finally, we add flow relations: $F \leftarrow F \cup (T_{21},P_{17})$, $F \leftarrow F \cup (P_{17},T_{22})$ et $F \leftarrow F \cup (T_{22},P_o)$.

The changes made to render the CPr workflow opaque are depicted in Figure \ref{fig:prive_opacification}.
\input{opacification_prive.tex}

%% file: sogalgorithm.tex
	
\scalebox{0.87}{
	\begin{minipage}[t]{6.1cm}
		\null 
		\begin{algorithm}[H]
			\label{alg:OSOG}
			\caption{SOG-based Opacification}
			%\tiny
			\scriptsize 
			\SetAlgoLined
			\SetKwInput{KwFunc}{Procedure}
			\KwFunc{SOG-based Opacification ($(P, T, F,W),m_o,m_S,\Sigma_{o}\cup\Sigma_{u}$)}
			$\textbf{Vertices}\; V; \; \textbf{Edges}\; E$\;
			$\textbf{Aggregate} \; a,a'$\;
			$\textbf{Stack} \; st, \;CounterExample$\;
			$\textbf{Incidence Matrix} \; C$\;
			\Begin{
				$(Q,q_{init},\Sigma,\delta) \leftarrow \textbf{BuildReachabilityGraph}(P, T, F,W,m_o)$\;
				$S \leftarrow m_S$\;
				$a \leftarrow Saturate(\{q_{init}\})$\;
				
				\If{($a \subseteq S$)}{
					$CounterExample.\textbf{Push}(\epsilon,a,\epsilon,a)$;
				}
				%else if1
				$V \leftarrow {a}; E \leftarrow \varnothing $\;
				$trace\leftarrow\emptyset$\;
				$st.\textbf{push}\left((a,EnableObs(a))\right)$\;
				\While{$(st \neq \emptyset)$}{
					$(a,enb) \leftarrow st.\textbf{Top}()$\;
					\eIf{$(enb \neq \emptyset)$}{
						$st.\textbf{Pop}()$\;
					} %else if2
					{
						$t \leftarrow RemoveLast(st.Top.Second())$\;
						$a' \leftarrow Img(a,t)$\;
						$a' \leftarrow Saturate(a')$\;	
						\eIf{(Treated(a'))}{
							$E \leftarrow E \cup t$\;
							$Save(a \xrightarrow{t} a')$\;		
						} %else if3
						{
							\If{($a' \subseteq S$)}{
								$Trace= Print$\\ \nonl $CounterExample()$\;
								$CounterExample.$\\ \nonl$\textbf{Push}(trace,a,t,a')$\;	
							} %end if4
							
							$V \leftarrow V \cup \{a'\}$\;
							$ E \leftarrow E \cup t$\;
							$Save(a \xrightarrow{t} a')$\;
							$st.\textbf{Push}(a',EnableObs(a'))$\;
							
						}%end if3
					} %end if2
				} %While
				%end if1
				\If{$(CounterExample \neq \emptyset )$}{$\textbf{Opacification}()$\;}
			} % endbegin
		\end{algorithm}
	\end{minipage}%
}
%\vspace{5cm}
\scalebox{0.93}{
	\begin{minipage}[t]{6.1cm}
		\null
		\begin{algorithm}[H]
			\label{alg:Opacification}
			\caption{Opacification}
			%\tiny
			\scriptsize 
			\SetAlgoLined
			\SetKwInput{KwFunc}{Procedure}
			\KwFunc{Opacification$()$}
			\Begin{
				$minSL=\textbf{ComputationMinSL}(L()$\;
				\While{$(CounterExample \neq \emptyset)$}{
					$(trace,a,t,a')\leftarrow CounterExample.\textbf{Top}()$\;
					\If{($NotTreated(a'))$}{
						\ForEach{$u \;\textbf{in }\; minSL$}{
							\If{$(u = trace)$}{
								\tcc{SOG Opacification}
								$q_{new} = new \; State()$\;
								$a' \leftarrow a' \cup \{q_{new}\}$\;
								$Save(a \xrightarrow{t}a')$\;
								\tcc{LTS Opacification}
								$q\leftarrow CounterExample.\textbf{Top.Fourth}()$\;
								$t_{new} \leftarrow new \; UnobservableTransition()$\;
								$Q \leftarrow Q \cup {q_{new}}$\;
								$\Sigma_{u} \leftarrow \Sigma_{u} \cup t_{new}$\;
								$\delta(q,t_{new})=q_{new}$\;
								\tcc{Petri net Opacification}
								$p_{new} \leftarrow new Place()$\;
								$P \leftarrow P \cup p_{new}$\;
								$T \leftarrow T \cup t_{new}$\;
								$p \leftarrow$ $\textbf{getPlace}()$\;
								$F \leftarrow F \cup (p, t_{new})$\;
								$F \leftarrow F \cup (t_{new}, p_{new})$\;
								$W \leftarrow W \cup \{ ((p, t_{new}) \longmapsto 1) , ((t_{new}, p_{new}) \longmapsto 1)  \}$\;
								$C(p_{new},t_{new}) \leftarrow W(t_{new},p_{new}) - W(p_{new},_{tnew}) $\;
							}	%endIf1
						}%end foreach	
					}%end if not treaed
					$CounterExample.\textbf{Pop}()$\;
				}%end while
			}
		\end{algorithm}
	\end{minipage}
}

%% file: opacification_prive.tex
\begin{figure*}[!htbp]
\begin{minipage}{0.2\textwidth}
	 \begin{subfigure}[SOG]{
			\centering
			\includegraphics[width=3in]{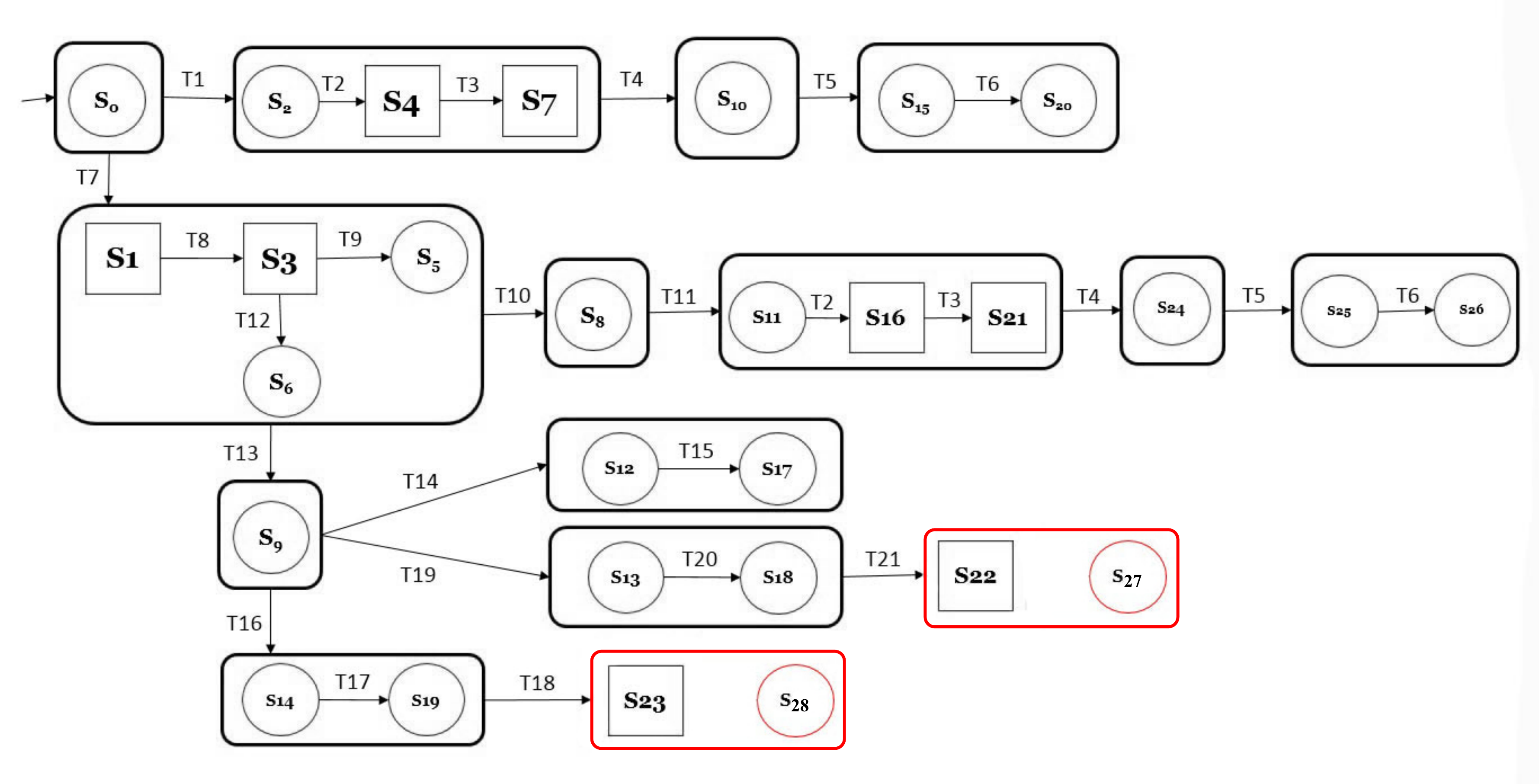}
		}
	\end{subfigure}
	\hfill
	\begin{subfigure}[LTS]{
			\centering
			\includegraphics[width=3in]{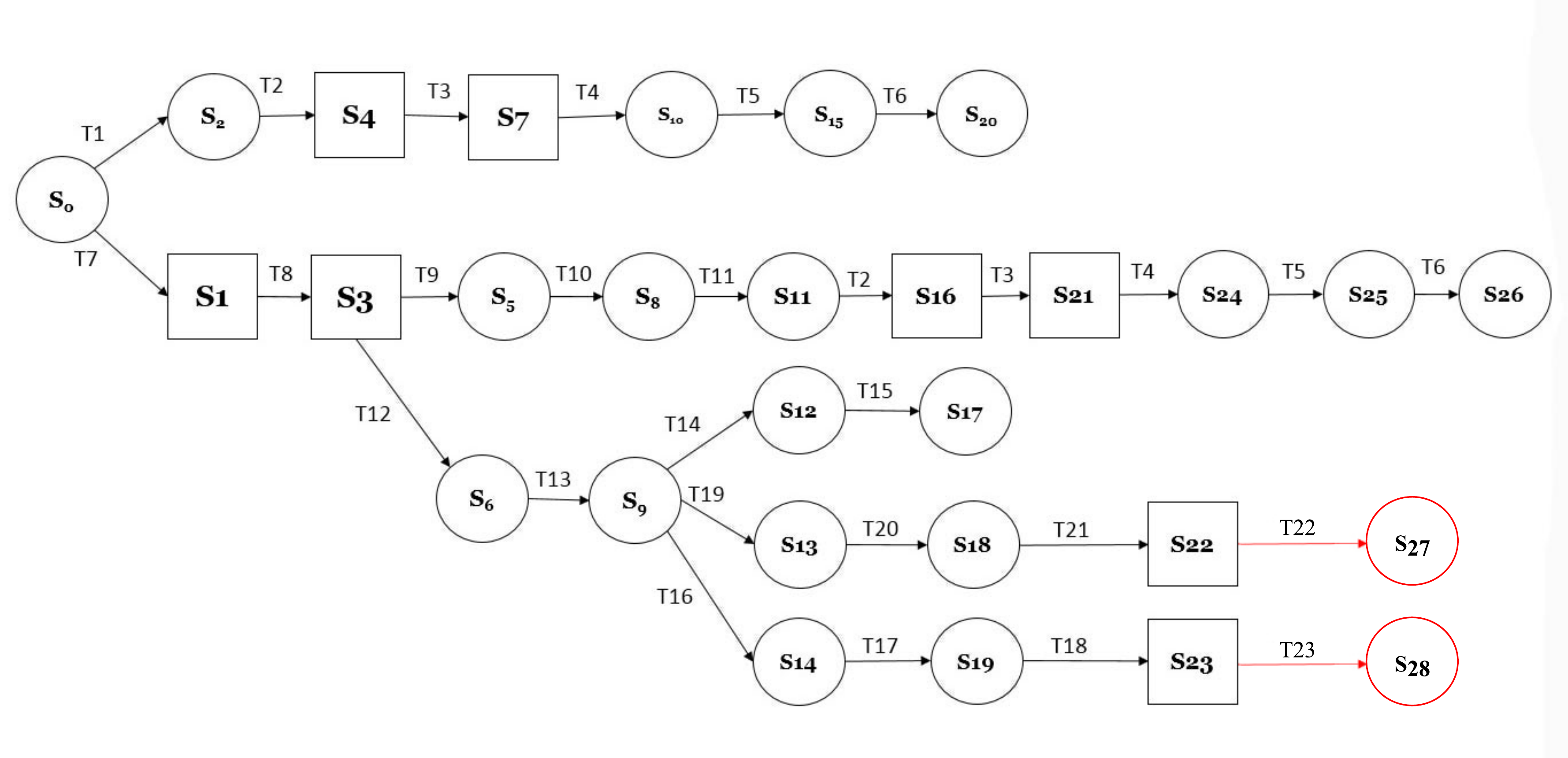}
		}
	\end{subfigure}
\end{minipage}
\hspace{5cm}
\begin{minipage}{0.2\textwidth}
\begin{subfigure}[oWF-net]{
		\centering
		\includegraphics[width=2in]{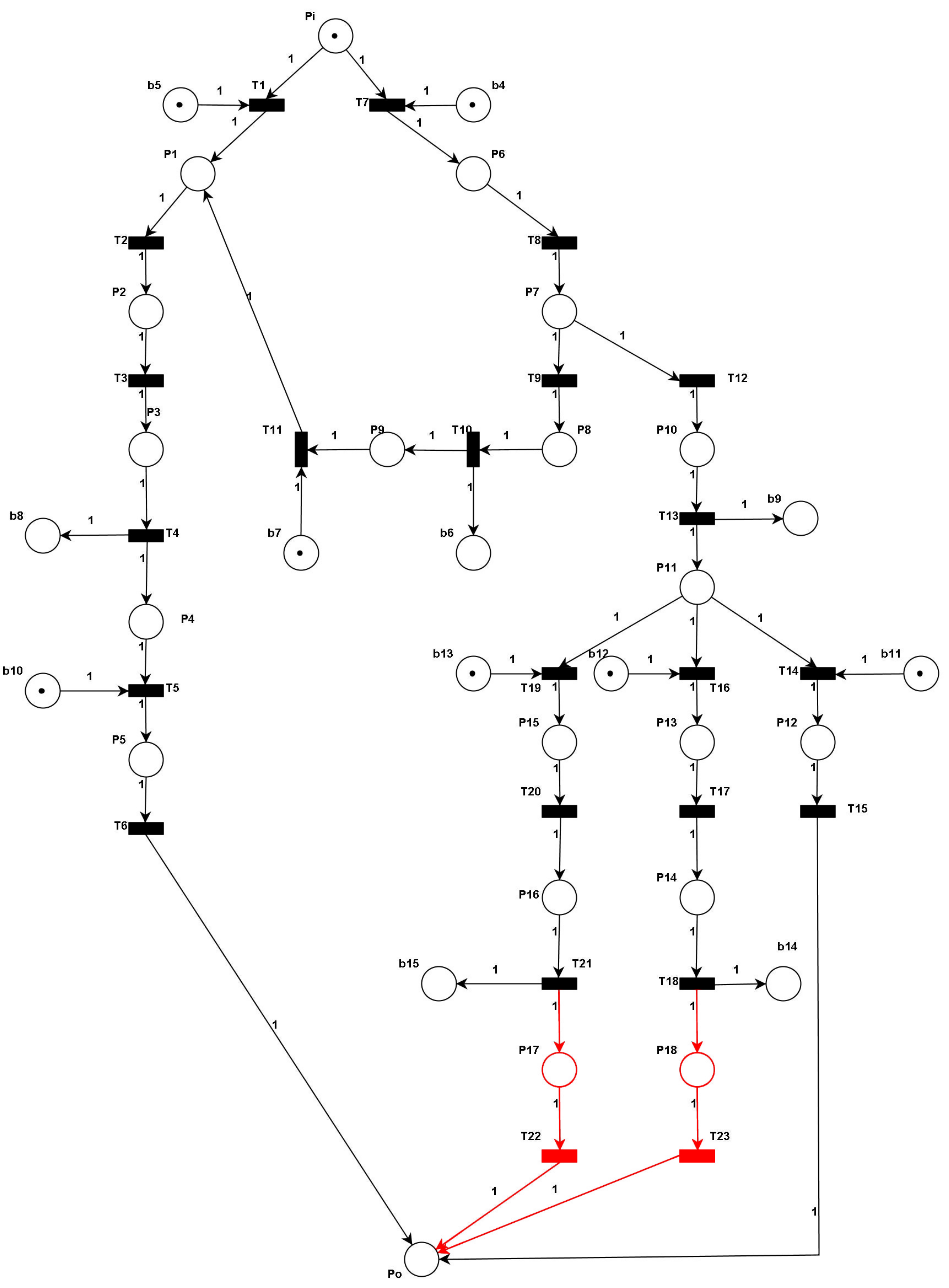}
	}
\end{subfigure}
\end{minipage}
	\caption{The CPr workflow opacification}
\label{fig:prive_opacification}
\end{figure*}

%% file: conc.tex
\addcontentsline{toc}{section}{Conclusion \& Perspectives}
\section{Conclusion  \& Perspectives}
In this paper, we used opacity, a generalization of many security properties, as a means to track the information flow in an IoT-based medical application. We introduced a model to analyze the behavior of an IoT-based heart attack detection system discussing how an observer may infer personal patient information. Our work aims at detecting security leaks in our system, using SOG-based algorithms for the on-the-fly verification of opacity variants (simple, $K$-step weak, and $K$-step strong opacity). We have also proposed a novel, SOG-based approach for opacity enforcement of Petri net-modeled systems. The main contribution of this work is to propose an efficient algorithm for enforcing simple opacity by padding the system with minimal dummy behavior. In our future research, we will explore the same idea of enforcement for other opacity variants such as $K$-step weak and $K$-step strong opacity. Furthermore, we also hope to extend this work to take into account different types of enforcement, such as supervisory control for opacity and finding the supremal sub-language, instead of computing the minimal super-language.